\documentclass[12pt]{article}
\usepackage{a4}
\usepackage{bbold}

\newcommand{\eqn}[1]{(\ref{#1})}
\def\uni{\mathbb{1}}
\def\bea{\begin{eqnarray}}
\def\eea{\end{eqnarray}}
\def\be{\begin{equation}}
\def\ee{\end{equation}}
\def\bei{\begin{itemize}}
\def\eei{\end{itemize}}
\def\bem{\begin{minipage}[c]}
\def\eem{\end{minipage}}

\def\nn{\nonumber}
\def\d{\delta}

\def\g{\gamma}
\def\tS{\tilde S}
\def\p{\partial}

\newcommand{\XX}[2]{\{X^#1,X^#2\}}
\newcommand{\bX}{\hat{\bf X}}
\newcommand{\ft}[2]{\textstyle\frac{#1}{#2}}

\newcommand{\ed}{\end{document}}

\newcommand{\NPB}[3]{{Nucl.\ Phys.} {\bf B#1} (#2) #3}
\newcommand{\IJMPA}[3]{{Int.\ J.\ Mod.\ Phys.} {\bf A#1} (#2) #3}
\newcommand{\CMP}[3]{{Commun.\ Math.\ Phys.} {\bf #1} (#2) #3}

\newcommand{\PLB}[3]{{Phys.\ Lett.} {\bf B#1} (#2) #3}
\newcommand{\JHEP}[3]{{JHEP} {\bf #1} (#2) #3}

\newcommand{\T}[1]{``{\em #1}''} 
\newcommand{\hepth}[1]{{\tt hep-th/#1}}

\begin{document}

\thispagestyle{empty}
\begin{flushright}
{\small hep-th/0009193}\\
{\small AEI-2000-056}\\[3mm]
\end{flushright}

\vspace{1cm}
\setcounter{footnote}{0}
\begin{center}
{\Large{\bf Vertex Operators for the Supermembrane and Background 
Field Matrix Theory}
    }\\[14mm]

 {\sc Jan Plefka\footnote{Talk given 
at the {\it E.S. Fradkin Memorial Conference,
Moscow 2000,}
and {\it Strings 2000, University of Michigan, Ann-Arbor.}}}\\[10mm]

{\em Albert-Einstein-Institut}\\
{\em Max-Planck-Institut f\"ur
Gravitationsphysik}\\
{\em Am M\"uhlenberg 1, D-14476 Golm, Germany}\\
{\footnotesize \tt plefka@aei-potsdam.mpg.de}\\[7mm]

{\sc Abstract}\\
\end{center}
We derive the vertex operators that are expected to govern the emission 
of the massless $d=11$ supermultiplet from the supermembrane in the
light cone gauge.  Our results immediately imply the  
linear coupling of matrix theory to an arbitrary supergravity background
to all orders in anticommuting coordinates. Finally we address the
definition of n-point tree level and one-loop scattering amplitudes.
The resulting 3-point tree level amplitudes turn out to agree 
with $d=11$ supergravity and are completely fixed by supersymmetry
and the existence of a normalizable ground state.

\vfill
\leftline{{\sc September 2000}}

\newpage
\setcounter{page}{1}
\section{Introduction}
One of the most pressing question in string theory to date is as to
what the precise microscopic degrees of freedom of M theory are.
On the face of it the 11 dimensional supermembrane \cite{BST} appears as
a natural candidate for M theory, as it sits atop of the main contenders
for a unified theory of quantum gravity: supergravity \cite{CJS}, superstrings
\cite{superstring} and
matrix theory \cite{BFSS}, 
which all are obtained through certain limits of the membrane
model. This is obvious for the case of
eleven dimensional supergravity, where one simply discards
all internal excitations of the membrane, being left with a first
quantized description of supergravity in form of the 11 dimensional
 superparticle.
Similarly one reaches type IIA superstrings at the kinematic level
through a procedure called double-dimensional reduction \cite{DHIS}. 
And finally matrix theory, a proposed candidate 
for light cone M theory, emerges as
a finite N regularization of the light cone supermembrane.
Despite these features the study of the fundamental supermembrane 
has not received much attention caused by the tremendous 
difficulties one encounters once one
turns to a quantization of the model. Opposed to the particle and the
string the membrane is a nonlinear interacting field theory, with
a priori no well defined perturbative scheme in form of a sum over worldvolumes.
Moreover, the model possesses a continuous spectrum, which, following the
insights from the matrix theory proposal, should be interpreted as
a second quantized feature. In order to make progress we believe that 
a starting point for a quantum treatment of the supermembrane should
be to clarify what the sensible quantities or operators are 
whose expectation values one would like to compute. We here
want to push forward the concept of membrane vertex operators, which 
are expected to govern the emission of the massless d=11 supermultiplet 
from the  membrane worldvolume. With these we are able to {\it define}
scattering amplitudes in membrane theory, thus making the supermembrane
more ``computable''. Moreover, these operators immediately translate
into the corresponding objects in matrix theory, thus yielding the
linear order background field coupling of matrix theory to all
orders in fermions. We shall show that the
resulting 3-point amplitudes agree with d=11 supergravity, and
comment at ongoing work to compute the membrane four graviton amplitude,
which yields the famous ${\cal R}^4$ quantum correction to the d=11
supergravity action.

In analogy to superparticle and superstring theory the supermembrane
vertex operators are naturally defined as the linear coupling of the
supergravity background fields $g_{\mu\nu},C_{\mu\nu\rho}$ and
$\psi^\alpha_\mu$ to the embedding coordinates $X^\mu(\xi)$ 
and $\theta_\alpha(\xi)$, where $\xi$ parameterizes the three dimensional
membrane world-volume.
In principle these operators are deducible from the known
background field action of the supermembrane in superspace \cite{BST}
\be
S=\int d^3\xi  \sqrt{-g[Z(\xi)]}
+\ft 1 6 \epsilon^{ijk}\, \pi_i^A\pi_j^B\pi_k^C\, B_{CBA}[Z(\xi)]\, 
\label{superspace}
\ee
where $E^A_M$ denotes the super-dreiundvierzig-bein, $B_{CBA}$ the 
super three-form and we have $Z^A=(X^\mu(\xi),\theta^\alpha(\xi))$, 
$\pi_i^A=\frac{\partial Z^M}{\partial \xi^i}\, E_M^A$ and
$g_{ij}=\pi_i^r\,\pi_j^s\,\eta_{rs}$. Obtaining the linearized action 
in components from \eqn{superspace}, however, is a highly nontrivial task. 
To date this
has only been achieved up to second order in anticommuting coordinates
$\theta$ \cite{dWPP} through a process called gauge completion \cite{CF}, i.e.
\bea
E_\mu^r&=& {e_\mu^r} + 2\,\bar\theta\Gamma^r{\psi_\mu} 
+ \bar\theta\Gamma^r\Bigl [ -\ft 14 \,\hat\omega_\mu\cdot \Gamma + T_\mu
\cdot \hat F\Bigr]\theta + \ldots + {O}(\theta^{32})
\nn\\
E_\mu^\alpha&=& {\psi_\mu^a}-\ft 1 4 \,{\hat\omega_\mu{}^{rs}}
\, (\Gamma_{rs}\theta)^\alpha
+(T_\mu\cdot \hat F\, \theta)^\alpha
+\ldots + {O}(\theta^{32}) \nn \\
B_{\mu\nu\rho} &=& {C_{\mu\nu\rho}} -6\,\bar\theta\Gamma_{[\mu\nu}
 {\psi_{\rho]}}
-3\, \bar\theta\Gamma_{[\mu\nu}\Bigl[-\ft 14 \,\hat\omega_{\mu]}
\cdot \Gamma + T_{\mu]}
\cdot \hat F\Bigr]\theta
\ldots + {O}(\theta^{32})\nn
\eea
Note that these expansions in principle extend all the way up to order 32
in $\theta$'s. We hence see that obtaining the complete expressions
for the {\it covariant} vertex operators via superspace seems to be
beyond reach\footnote{There exists an alternative and potentially 
more effective
procedure via a normal coordinate expansion in superspace 
\cite{Grisaru}, which has so far not been applied to the eleven dimensional
case.}. This motivates the transition
to a light cone description of the supermembrane in general backgrounds
for which one expects a substantial degree of
simplification, as may be seen already the level of the flat space action.
We shall see that here the $\theta$ expansion terminates at order four.

Imposing the light cone condition $X^+=p^+\, \tau$ and $\Gamma^+\Theta=0$ 
by $\kappa$-symmetry the flat space action of the supermembrane reads \cite{dWHN}
\be
{
{\cal L}}= \ft 1 2 (DX^a)^2  - \ft 14 \{X^a,X^b\}^2
- i \theta\, D\theta - i \theta\, \gamma_a \{ X^a ,\theta\} 
\label{action}
\ee
where $X^a=X^a(\tau,\sigma_1,\sigma_2)$ and
$\theta^\alpha=\theta^\alpha(\tau,\sigma_1,\sigma_2)$ 
denote the transverse ($a=1,\ldots 9$, $\alpha=1,\ldots, 16$) coordinates.
We have used the covariant derivative $D X^a:= \partial_\tau X^a- 
\{\omega,X^a\}$ along with the Poisson bracket   
$\{ A, B\} := \epsilon^{ij}\, \partial_{\sigma_i}A\, \partial_{\sigma_j} B$.
The gauge field $\omega$ entering the covariant derivative $D$ stems from
a residual symmetry of area preserving diffeomorphisms \cite{GH}
\be
\delta {X^a}=\{\xi,{X^a}\},\qquad 
\delta {\theta}=\{\xi,{\theta}\},\qquad
\delta \omega=\partial_\tau+ \{\xi,\omega\}
\label{APD}
\ee
The action \eqn{action} is moreover invariant under the set of supersymmetry
transformations
\bea 
\delta X^a &=& -2\, {\epsilon}\,\gamma^a\theta \qquad \qquad
\delta\omega = -2\, {\epsilon}\,\theta \nn\\
\delta \theta &=& i \, DX^a\, \gamma_a\,{\epsilon}-\ft i 2
\{X^a,X^b\}\,\gamma_{ab}\, {{\epsilon}} + 
{{\eta}}
\label{susy}
\eea
with 32 components, 16 from $\eta$ and 16 from $\epsilon$, a remnant of the
11 dimensional origin of the model. Viewing the parameters $(\sigma_1,\sigma_2)$ 
as internal degrees of freedom the light cone supermembrane may be
understood as a supersymmetric quantum mechanical system equipped with 
an infinite dimensional gauge group of area preserving diffeomorphisms.

\section{Vertex Operator Construction}
The graviton, three form and gravitino emission operators that we seek to 
construct take the general form 
\bea
V_{{h}}&=& {h_{ab}\, } \int d\tau d^2\sigma 
\, {\cal O}^{{ab}}{[X^a(\tau,\sigma_i),
\theta(\tau,\sigma_i)]}\, e^{i {k}\cdot X +i k_+X^+}\nn\\
V_{{C}}&=& {C_{abc}\, } \int d\tau d^2\sigma 
\, {\cal O}^{{abc}}{[X^a,
\theta]}\, e^{i {k}\cdot X  +i k_+X^+}\nn\\
V_{{\psi}}&=& {\psi_a^\alpha} \int d\tau d^2\sigma 
\, {\cal O}^{{a}}_{{\alpha}}{[X^a,
\theta]}\, e^{i {k}\cdot X  +i k_+X^+}
\label{seek}
\eea
with the polarizations $(h_{ab},C_{abc},\psi_a^\alpha)$ and momenta
$k_a$ and $k_+$. The polarizations 
are subject to {on-shell constraints}, e.g. 
$k^a\, h_{ab}=h_{aa}=0=k^a\, C_{abc}$.\footnote{A derivation of this
may be found in \cite{GGK}. Note that in the above we are suppressing
all terms coupling to $k_-$, thus effectively setting $k_-=0$ in order
to decouple $X^-$ in the expressions, a complicated function in $X^a$ 
and $X^+$.
This is standard practice in light cone string theory, albeit questionable,
as with it all transverse momenta become complex. A cleaner way to state
this point is that one consistently ignores all terms coupling to
$k_-$ in the computations, assuming that they work out by themselves.}
The strategy for the explicit construction of the operators
${\cal O}^{ab}$, ${\cal O}^{abc}$ and ${\cal O}^a_\alpha$ is rather
simple. The (unknown) full background field action of the supermembrane
transforms covariantly under supersymmetry as
\be
{\d}{\cal L}_{{\mbox{\tiny
 full}}}[X,\theta; h,C,\psi]=
{\cal L}_{{\mbox{\tiny full}}}[X,\theta; {\hat\d}h,{\hat\d}C,{\hat\d}\psi]
\ee
where $\delta$ denotes the supersymmetry variation of $X^a,\theta$ and
$\omega$ of \eqn{susy}, whereas $\hat\d$ are the induced 
light-cone supergravity
variations of the background fields $h_{ab},C_{abc}$ and $\psi_a^\alpha$
whose precise expressions may be found in \cite{GGK,DNP}.
Hence the vertex operators, being the linear background field couplings,
must transform into each other under supersymmetry according to
\bea
{\d} V_h &=& V_{{\hat\d}\psi[h]} \nn \\
{\d} V_C &=& V_{{\hat\d}\psi[C]} \nn \\
{\d} V_\psi &=& V_{{\hat\d} h} + V_{{\hat\d} C} 
\eea
So for example under the linear part of the supersymmetry transformations
of \eqn{susy} ${\d}X^a=0$ and ${{\d}\theta={\eta}}$ 
the vertex operator $V_\psi$ transforms into a sum of the graviton
and three-form vertices, whose polarizations are then given by
${{\d}h_{ab}=-\tilde\psi_{(a}\gamma_{b)}{\eta}}$
and ${{\d}C_{abc}=\ft32
\tilde\psi_{[a}\gamma_{bc]}{\eta}}$ respectively, the linearized form
of the supergravity transformations parametrized by the same $\eta$
entering \eqn{susy}. 

It turns out that this requirement of covariance under supersymmetry
completely determines the form of \eqn{seek}. The results read
\cite{DNP}
\bea
V_h &=&
 {h_{ab}}\, \Bigl [ DX^a\, DX^b - {\{X^a,X^c\}\,\{X^b,X^c\}
 - i \theta\gamma^a\,\{X^b,\theta\} } \nn\\
&& -2 DX^a\, R^{bc}\, k_c {-6 \{X^a,X^c\}\, R^{bcd}\, k_d }
+ 2 R^{ac}\,R^{bd}\, k_c\, k_d \Bigr ]\,\nn\\
V_C &=&
-{C_{abc}}\, {DX^a\, \XX{b}{c}} +{F_{abcd}}\Bigl[
(DX^a-\ft 23 R^{ae}\, k_e)\, R^{bcd} \nn\\
&& {-\ft 12 \XX{a}{b}\, R^{cd} -\ft 1 {96}\, \XX ef\, \theta 
\gamma^{abcdef}
\theta }\Bigr ]\nn\\
V_\psi &=&
{\psi_a}\, \Bigl [ \, \Bigl (DX^a-2 R^{ab}\, k_b +\gamma_c\,
{\XX ca}\, \Bigr ) \theta \Bigr ] \nn\\[-0.1cm]
&&\!\!\!\!+ {\tilde\psi_a}\, \Bigl [ \gamma\cdot DX\,  \Bigl (DX^a- 2R^{ab}\, k_b
+\gamma_{c} {\XX ca} \Bigr )\,\theta \nn\\
&&\!\!\!\!+\ft 12 \gamma_{bc}\, {\XX b c} \, 
( DX^a- {\XX a d} \, \gamma^d\, )\theta\nn\\
&&\!\!\!\!+ 8  \gamma_b\theta\, {\XX b c }\, R^{cad}\, k_d 
  + \ft 5 3 \gamma_{bc}\theta\, {\XX b c} R^{ad}\,k_d\nn\\
&&\!\!\!\!+ \ft 4 3 \gamma_{bc}\theta\, \Bigl ( {\XX a b}\, 
R^{cd} + {\XX c d} R^{ab}
\Bigr ) k_d \nn\\
&&
+\ft 2 3 i \,\Bigl ( \gamma_b\theta\, {\{X^a,\theta\}}\gamma^b\theta
-\theta\, {\{X^a,\theta\}}\theta \,\Bigr ) \nn\\
&&\!\!\!\!+ \ft 8 9 \gamma^b\theta\, R^{ac}\, R^{bd}\, k_c k_d
\,\Bigr ]
\label{result}
\eea
where $R^{ab}=\ft 14 \theta\gamma^{ab}\theta$ and 
$R^{abc}=\ft 1{12} \theta\gamma^{abc}\theta$, suppressing the
overall $\exp(ik\cdot X)$ of \eqn{seek}. The vertices for 
$h_{+a},h_{++},C_{+ab},\psi_+$ and $\tilde\psi_+$ are also known;
$\psi$ and $\tilde\psi$ are the light-cone decompositions of
the gravitino \cite{GGK}.

These results are subject to three stringent consistency checks.
Firstly, the vertex operators are invariant under the following
background field symmetries: 
\bea
\d h_{ab}&=& k_{(a}\, \xi_{b)} \qquad \mbox{(coordinate transformations)}\nn\\
\d C_{abc} &=& k_{[a}\,\xi_{bc]} \qquad\mbox{(tensor gauge transformations)}\nn\\
\d \psi_a&=& k_a\epsilon     \quad\,    \qquad\mbox{(field independent SUSY)}\nn
\eea   
Secondly, the point particle limit of the membrane vertices \eqn{result},
which amounts to simply droping the terms involving the Poisson brackets
${\{., .\}}$ yields the
d=11 superparticle vertex operators of Green, Gutperle and Kwon \cite{GGK}.

Finally, a stronger check is to perform a double dimensional 
reduction \cite{DHIS}
of the
membrane vertices, which should reduce them to the type IIA superstring
vertex operators. This reduction procedure is performed by wrapping
the $\sigma_2$ coordinate of the membrane around a target space circle 
along $X^9$
\be
X^a(\tau,\sigma_1,\sigma_2) \rightarrow 
\left(\begin{array}{c}X^i(\tau,\sigma_1)\\
X^9=\sigma_2\end{array}\right)\quad i=1,\ldots,8
\label{dd1}
\ee
along with
\be
\theta_\alpha\rightarrow \left(\begin{array}{c}S_a(\tau,\sigma_1)\\
\tilde S_{\dot{a}}(\tau,\sigma_1)\end{array}\right)\quad a,\dot{a}=1,\ldots,8
\label{dd2}
\ee
Quite remarkably under this description the vertex operators of \eqn{result}
factorize into left and right moving contributions. Let us  
demonstrate this 
for the graviton vertex reduction. Under \eqn{dd1} and \eqn{dd2}
we have
$$
\XX ij =0 \qquad \XX i 9 = \partial_{\sigma_1}X^i\qquad \omega=0
$$
\be
\theta \gamma^{ij}\theta = S\Gamma^{ij}S + \tilde S\Gamma^{ij}\tilde S
\qquad
\theta {\gamma^{ij9}}\theta = S{\Gamma^{ij}}S 
- \tilde S{\Gamma^{ij}}\tilde S
\ee
where $\Gamma^i$ are the standard SO(8) $\Gamma$-matrices.
Then 
\bea
V_h|_{\mbox{\tiny {DDR}}} 
&\rightarrow& 
h_{ij} \Big[ \p_0 X^i \p_0 X^j -\p_1 X^i \p_1 X^j
    - \ft12 \p_0 X^i (S\Gamma^{jm} S \nn\\
&& + \tS \Gamma^{jm} \tS ) k_m  
 + \ft12 \p_1 X^i (S\Gamma^{jm} S - \tS \Gamma^{jm} \tS ) k_m \
   + \ft14 S \Gamma^{im} S \, \tS \Gamma^{jn} \tS k_m k_n \Big] \nn\\
&=&h_{ij} 
\Big( \p_+ X^i - \ft12 S\Gamma^{im} S k_m \Big)
\, \Big( \p_- X^j - \ft12 \tS\Gamma^{jn} \tS k_n \Big)\nn
\eea
which is nothing but the IIA graviton vertex.

\section{Matrix Theory in Background Fields}\
The obtained results may be directly translated to matrix theory, which
emerges from a supersymmetry preserving ``discretization'' of the membrane
spacesheet. For this one replaces the infinite dimensional 
gauge group of area preserving
diffeomorphisms by the large $N$ limit of SU($N$) \cite{GH}. This well known
prescription here amounts to the replacements
$$
X^a(\tau,\sigma^1,\sigma^2) \rightarrow {\bf X}^a_{mn}(\tau)\qquad
\theta^\alpha(\tau,\sigma^1,\sigma^2) \rightarrow {\bf \Theta}^\alpha_{mn}
(\tau)\qquad
m,n=1,\ldots, N$$
\be
\{ . , . \} \rightarrow i\, [ . , . ]\qquad\qquad
\ft{1}{4\pi} \int d^2\sigma (\ldots ) \rightarrow \ft 1 N\,
{\mbox{STr}[}\ldots {]}
\ee
$\mbox{STr}$ denotes the symmetrized trace whose introduction becomes
necessary due to ordering ambiguities 
for the composite operators that we have been discussing. 
The $\mbox{STr}$ prescription guarantees
that all manipulations performed in section 2 for the continuous membrane model go
through for the matrix model as well. In principle there may exist a 
less symmetric ordering that also works, but the differences
to the $\mbox{STr}$ prescription will be subleading in $N$.

Hence the weak (i.e. linear coupling) background field action of matrix theory is 
now known to {\it {all}} orders in ${\bf \Theta}$ and derivatives
$\partial/\partial {\bf X}^a$
\be
{\cal S}_{\mbox{\footnotesize MT}}= \int d\tau 
\Bigl ( {\cal L}_0 + V_{h(X)} + V_{C(X)} 
+ V_{\Psi(X)}\, \Bigr ) 
\ee
where e.g. the graviton coupling takes the form
\bea
V_{h(X)}&=&\mbox{STr}\Bigl [\Bigl\{\dot {\bf X}^a\dot {\bf X}^b 
+ [{\bf X}^a,{\bf X}^c]\,
[{\bf X}^b,{\bf X}^c]+
+{\bf \Theta}\gamma^a\,[{\bf X}^b,{\bf \Theta}] \nn\\
&&\quad -\ft 12 \dot{\bf X}^a\, ({\bf \Theta}\g^{bc}{\bf \Theta})\, 
\frac{\p}{\p{{\bf X}^c}} 
-\ft 12 i [{\bf X}^a,{\bf X}^c]\, ({\bf \Theta}\g^{bcd}{\bf \Theta})\, 
\frac{\p}{\p{{\bf X}^d}}\nn\\
&&\quad+ 2({\bf{\Theta}}\g^{ac}{\bf \Theta})\, ({\bf \Theta}\g^{bd}{\bf \Theta})
\frac{\p}{\p{{\bf X}^c}}\,\frac{\p}{\p{\bf X}^d}
\Bigr\}\, h_{ab}({\bf X})\Bigr ]
\eea
Note that we have performed a Fourier transformation of
\eqn{result} to target configuration space. These results agree with and go
beyond the explicit matrix theory current calculations of Taylor and V.Raamsdonk
\cite{TvR}, which so far have been performed up to order $o(\theta^2)$ 
and linear order in $\partial/\partial {\bf X}^a$.

\section{Scattering Amplitudes}
We now turn to the discussion of three point tree level scattering amplitudes.
For this it is advantageous to work in the framework of the 
finite $N$ matrix theory. In order to define a tree level amplitude we
first split off the center of mass degrees of freedom of the matrices by
writing
\be
{\bf X}^a={x^a}\, \uni + \hat{\bf X}^a \qquad
{\bf \Theta}^a={\theta^a}\, \uni + \hat{\bf \Theta}^a 
\label{zero}
\ee
with traceless matrices $\hat{\bf X}^a$ and $\hat{\bf \Theta}$. An 
asymptotic 1-graviton state in matrix theory is then given by
\be
|\mbox{IN}\rangle\rangle= 
|k_1,h_1\rangle_{{x,\theta}} \otimes 
|\mbox{GS}\rangle_{\hat{\bf X},\hat\Theta}
^{\mbox{\tiny SU(N)}}
\ee
where $|k_1,h_1\rangle$ is the graviton state of the superparticle
\cite{PW,GGK} and 
$|\mbox{GS}\rangle$ denotes the {\it exact} SU($N$) normalized zero energy
groundstate, whose explicit form is unknown but is known to exist \cite{SS}. 
The tree level three point amplitude is then defined by
\be
{\cal A}_{\mbox{\tiny 3-point}}=\langle\langle \,1\,|\, V_2\,|\,3\,\rangle\rangle
\label{3pt}
\ee
where one inserts the graviton vertex operator
\be
V_2=h^{(2)}_{ab}\, \ft 1 N \, \mbox{STr}\Bigl[({p^a}\,{p^b}  + 2\, {p^a}\, 
\hat{\bf P}^b
+ \hat{\bf P}^a\,\hat{\bf P}^b+ [\bX^a,\bX^c]\, [\bX^b,\bX^c]\, )\,
e^{ik\cdot\bX}\Bigr]\, e^{ik\cdot{x}}
+ \mbox{\small fermions}
\label{V2}
\ee
One may wonder how one could ever evaluate \eqn{3pt} upon inserting
\eqn{V2} without the knowledge
of $|\mbox{GS}\rangle$. The first contribution to \eqn{3pt} takes the
form
\be
\langle k_1,h_1| {p^a}\,{p^b}e^{ik\cdot{x}}|k_3,h_3\rangle\,
h^{(2)}_{ab}\,  
\langle\mbox{GS}|\mbox{STr}\, e^{ik\cdot\bX}|\mbox{GS}\rangle
\label{kh}
\ee
Now by SO(9) covariance 
$\langle\mbox{GS}|\mbox{STr}\, e^{ik\cdot\bX}|\mbox{GS}\rangle=N$, as
the only SO(9) scalar it could depend on would be $k^2$ which vanishes
on shell. It must then be a constant which is fixed to be $N$ by considering the
$k^a\rightarrow 0$ limit. Remarkably the remaining two terms in \eqn{3pt}
upon inserting \eqn{V2} vanish by a combination of SO(9) covariance
and on-shell arguments:
\be
h_{ab} \, \langle\mbox{GS}|\mbox{STr}\, \hat{\bf P^b} \, e^{ik\cdot\bX}
|\mbox{GS}\rangle \sim k^b\, h_{ab}=0
\ee
$$
h_{ab} \, \langle\mbox{GS}|\mbox{STr}\Bigl[( \hat{\bf P^a}\, \hat{\bf P^b} 
+ [\bX^a,\bX^c]\, [\bX^b,\bX^c]\, )
\, e^{ik\cdot\bX}\Bigl ]
|\mbox{GS}\rangle \sim (k^a\, k^b+ c\, \delta^{ab})\, h_{ab}=0
$$
But as the first correlator in \eqn{kh} is nothing but the bosnonic
contribution to the 3-point
d=11 {\it superparticle} amplitude \cite{GGK} and as the fermionic terms
work out in a similar fashion we see that our 3-point tree level
amplitude 
\be 
\langle\langle \,1\,|\, V_2\,|\,3\,\rangle\rangle= 
\langle \,1\,|\, V_2\,|\,3\,\rangle_{{x,\theta}}\, \langle
\mbox{GS}|\mbox{GS}\rangle
\ee
agrees with the 3-point amplitude of d=11 supergravity!

Clearly the next step would be to study $n$-point tree level amplitudes
which should be given by
\be
{\cal A}_{\mbox{\tiny n-point}}
=\langle\langle \,1\,|\, V_2\, \Delta\, V_3\, \Delta \ldots \Delta\, V_{n-1}
|\,n\,\rangle\rangle
\label{npt}
\ee
where $\Delta$ denotes the propagator $1/(\ft 1 2\, {p_0}^2 + \hat{\bf H})$
built from the {\it interacting} membrane Hamiltonian $\hat{\bf H}$.
However, now we expect the details of the groundstate $|\mbox{GS}\rangle$
to enter the computation. Developing some perturbative scheme for
calculating \eqn{npt} would be highly desirable, but is conceivably
very complicated as it must involve an expansion in both the propagator
$\Delta$ and the groundstate $|\mbox{GS}\rangle$. 

Instead we shall briefly comment on ongoing attempts to compute loop
amplitudes within this scenario. Here, led by the formalism in
light cone superstring and superparticle theory, we propose to
define a membrane  $n$-point one-loop amplitude by the expression
\be
{\cal A}_{\mbox{ \tiny 1-loop, n-point}}
=\int d^{11}{p_0}\, \mbox{\bf Tr} ( \Delta\, V_1 \,
\Delta\, V_2\,  \Delta\, V_3\,  \Delta\, V_4\,  \ldots \Delta \, V_n)
\label{1lp}
\ee
where the trace is over the Hilbert space of $\hat{\bf H}$. Again
this appears as a daunting task, however the zero mode sector of
\eqn{1lp} already yields some amount of information. In particular
the trace over the fermionic zero mode $\theta$ of \eqn{zero}
tells us that all 2 and 3 particle amplitudes vanish at one loop,
as at least four vertex operators ($\leq$ 16 $\theta$'s) 
are needed to saturate the
fermion zero mode trace
\be
\mbox{\bf Tr} (\theta_{\alpha_1}\ldots
\theta_{\alpha_{{N}}})_{{\theta}} = \delta_{{N},16}\,
\epsilon^{\alpha_1\ldots\alpha_{16}}
\ee
In the pure graviton sector the first non-vanishing amplitude is
then the 4-graviton amplitude whose leading momentum dependence
is given by
\be
{\cal A}_{4h} = \epsilon^{\alpha_1\ldots\alpha_{16}}\, 
\gamma_{\alpha_1\alpha_2}^{a_1\, a_2}\ldots 
\gamma_{\alpha_{15}\alpha_{16}}^{a_{15}\, a_{16}}\,\,
R^{(1)}_{a_1a_2a_3a_4}\ldots R^{(4)}_{a_{13}a_{14}a_{15}a_{16}}
\int d^{11}\, {p_0}\, \mbox{\bf Tr}^\prime\, \Delta^4 \, .
\ee
We thus see the emergence of the expected ${\cal R}^4$ term 
\cite{GGV} in the kinematical sector, but it remains to
be seen what can be said about the remaining trace. Potentially
BPS arguments could here come to ones aid \cite{dWL}.

\section{Outlook}
In this talk we have constructed the supermembrane vertex operators
in the light cone gauge, which hopefully provide us with a new
tool in the study of quantum M theory. We have demonstrated their
reduction to the corresponding vertices of the d=11 superparticle
and d=10 type IIA strings. Moreover they yield
complete weak background field matrix theory action. It would
be interesting to clarify their relevance for the related
IKKT matrix model \cite{IKKT} and matrix string theory \cite{DVV}.
N-point tree level and 1-loop amplitudes were defined for the
model and
it was argued that the resulting 3-point tree level amplitudes
agree with d=11 supergravity.

Clearly there remains a host of open questions. On the conceptual
side one may ask where the multiparticle interpretation
of the membrane/matrix theory emerges in the outlined formalism.
After all our construction proceeds in complete analogy with
the first quantized particle and string theories. On the technical
side progress is clearly necessary for the evaluation of higher
loop amplitudes, both at tree and one loop level.

\vspace{6mm}
\noindent
{\bf Acknowledgement}

The material presented in this talk was obtained in collaboration
with A. Dasgupta and H. Nicolai \cite{DNP}. 
JP wishes to thank the organizers
of the E.S. Fradkin and Strings 2000 conferences for organizing a 
stimulating meeting and giving him the opportunity to present these results.

\ed